\documentclass[conference]{IEEEtran}
\IEEEoverridecommandlockouts
\usepackage{cite}
\usepackage{amsmath,amssymb,amsfonts}
\usepackage{algorithm}
\usepackage{algpseudocode}
\usepackage{graphicx}
\usepackage{textcomp}
\usepackage{xcolor}
\def\BibTeX{{\rm B\kern-.05em{\sc i\kern-.025em b}\kern-.08em
    T\kern-.1667em\lower.7ex\hbox{E}\kern-.125emX}}
\begin{document}

\title{Multi Armed Bandit Algorithms Based Virtual Machine Allocation Policy for Security in Multi-Tenant Distributed Systems.\\}


\author{\IEEEauthorblockN{1\textsuperscript{st} Pravin Patil}
\IEEEauthorblockA{\textit{Department of Computer Engineering,} \\
\textit{Assistant Professor.}\\
Pune Institute of Computer Technology. \\
E-mail: prpatil@pict.edu}
\and
\IEEEauthorblockN{2\textsuperscript{nd} Dr. Geetanjali Kale}
\IEEEauthorblockA{\textit{Department of Computer Engineering,} \\
\textit{Head \& Associate Professor.}\\
Pune Institute of Computer Technology. \\
E-mail: gvkale@pict.edu}
\and
\IEEEauthorblockN{3\textsuperscript{rd} Tanmay Karmarkar}
\IEEEauthorblockA{\textit{Department of Computer Engineering,} \\
\textit{Pune Institute of Computer Technology.}\\
E-mail: tanmaykarmarkar49@gmail.com}
\and
\IEEEauthorblockN{4\textsuperscript{th} Ruturaj Ghatage}
\IEEEauthorblockA{\textit{Department of Computer Engineering,} \\
\textit{Pune Institute of Computer Technology.}\\
E-mail: ruturajghatage33@gmail.com}
}

\maketitle

\begin{abstract}
Virtual machine allocation policies focus more on load balancing and power consumption aspects. Less attention is paid towards the security aspect while designing virtual machine allocation policy.  This work proposes a secure and dynamic VM allocation strategy for the multi-tenant distributed systems. The evaluation of the methodology demonstrates that the suggested strategy for placing virtual machines using the Thompson sampling approach is both effective and secure when compared to allocating virtual machines using the epsilon-greedy and upper confidence bound methods. This is evident in its lower levels of regret. Initially the cloud service provider was assigned with the task of allocating a VM to the client. This process was completely static. But the attacks can occur any time, any day and the attack can be of any type. We don't have control over that. This is why the historical data of Virtual machines was collected and studied to understand what their response to an attack was. If the attack was unsuccessful, rewards were granted to the tenant or virtual machine. If the attack was successful, rewards of the tenant or virtual machine were reduced which gave rise to an increase in regret. So now data was used to check if the machine is reliable or not and similarly it was allocated, giving rise to a dynamic system for VM allocation. 
The paper introduces an innovative approach of Multi Arm Bandit based Virtual Machine (VM) allocation policy. The authors train a model on a dataset that comprises known attacks and non-attacks, using a Weighted Average Ensemble Learning algorithm to enhance the f1 score. This ensemble algorithm assigns weights to each model to optimize its performance, providing superior results compared to traditional algorithms such as Logistic Regression, SVM, K Nearest Neighbors, and XGBoost. To detect suspicious activity, the authors propose a Stacked Anomaly Detector algorithm, which is trained on known non-attacks. This approach outperforms existing methods such as Isolation Forest 1, Isolation Forest 2, PCA 1, PCA 2, and Histogram Based Outlier Score. Overall, this paper offers a sophisticated and effective solution for VM allocation policies that can enhance the security of cloud-based systems.
\end{abstract}

\begin{IEEEkeywords}
Virtual Machine Allocation Policy, Multi Arm Bandit Algorithms, Regrets, Rewards, Anomaly detection, Multi-Tenant Distributed Systems.
\end{IEEEkeywords}

\section{Introduction}

In the context of VM allocation policy, we want to predict a sequence of allocations, where our only information about the allocations comes from recommendations of a set of previous VM allocations based on anomaly score [2, 19]. Note here that one of the measures to tackle VM is to get the VM anomaly score based-on various system parameters and decide the VM to be allocated. However, such an approach might not be feasible on two counts: (1) getting frequent anomaly score of a large pool of VM is computationally expensive, and (2) even the malicious nature of VMs can vary across the time i.e., the representation of malicious signatures varies over the time. Moreover, some of them, but not all, of these VMs might be unreliable, even adversarial. Traditional learning mechanisms might not capture all these characteristics dynamically. In such a scenario the paradigm of learning – multi arm bandits comes handy when the VMs need to face the unknown situation and absorb the malicious behaviour.

\section{RELATED WORKS}

\quad Currently, there is a significant increase in the complexity and rapid growth of network appliance services. Therefore, it is imperative to consolidate and integrate IT infrastructure to achieve centralized control and administration, leading to reduced ownership costs. In this context, cloud computing has emerged as a scientific concept that offers adaptable and expandable infrastructure and services to various entities. Cloud data is kept and retrieved on dedicated servers, with major cloud service providers like Amazon, Google's Application, IBM, and others serving as the customary models for computing and storage services, catering to individuals, businesses, and government initiatives [1,3]. Cloud computing has become popular, replacing autonomous computing, grid computing, and service computing.

Both physical and logical security challenges across various service models, such as software, platform, and infrastructure are addressed in Cloud Computing Security. Sharing resources in the cloud can increase the risk of security breaches as customers are exposed to other tenants with whom they share resources \cite{b4}. This sharing can potentially allow malicious tenants to attack other tenants on the same physical node, using side-channel attacks to access CPU, memory, and network patterns of other tenants and gain access to their private financial and economic data.

To defend against these attacks, multiple techniques have been developed, including the use of VM placement algorithms or VM allocation policies to reduce the possibility of co-location. To bolster security further, it's essential to implement a secure protocol that increases the difficulty for potential intruders trying to achieve co-residence and reduces the risk of side-channel attacks. To tackle the scalability problems in the existing algorithms this paper proposes a dynamic VM Placement algorithm.

The paper proposes a dynamic approach for allocating virtual machines (VMs) in server domains to minimize co-resistance among the machines, which in turn helps mitigate side channel attacks. This approach uses the baseline greedy algorithm and previously-selected-server-first policy (PSSF) to achieve its objectives. To tackle scalability challenges, the PSSF policy is utilized. This policy narrows down the search space by focusing on crucial actions that counter co-resistance attacks. The dynamic VM placement algorithm introduced aims to efficiently counter side-channel attacks.

Recently, many studies have proposed isolation techniques to reduce unpredictability and eliminate resource interference. Relevant research has been conducted on cloud vulnerabilities, such as detecting VM placement and side-channel attacks.

For example, \cite{b5} has proposed using virtualization to differentiate between legitimate and malicious customers who co-locate in the cloud. The study demonstrates how side-channel attacks can be used to extract cryptographic keys from victim virtual machines. To prevent this, the target virtual machine needs to be monitored frequently to observe its I-cache movements in detail, while filtering out noise from hardware and software influence and core immigration. A previously-selected-server-first-based scalable VM placement algorithm has been developed to mitigate side-channel attacks and make it difficult for attackers to obtain sensitive information from victim virtual machines.

A model called Utility-based Virtual Cloud Resource Allocation Model is proposed in [6] for allocating virtual cloud resources. This model considers the trade-offs between data center efficiency and application performance, and maximizes the utility based on meeting user requirements. The model includes a native decision process and a general decision process for resolving the allocation issue. This approach can enhance data center services when compared with other models. However, as the size of the cloud computing environment increases, there may be some issues with the model, such as performance delays.

In \cite{b7}, it is suggested that using side-channel attacks, it is possible to easily obtain confidential data from a machine. A combination of simulated firewall appliance and arbitrary encryption decryption is suggested to provide security against side-channel attacks in cloud computing because it ensures Reliability, Availability, and Security (RAS) and offers protection on all fronts. The proposed approach uses a virtual firewall in cloud servers to prevent the instantiation of virtual machines in the targeted VM during a side-channel attack. The approach also applies arbitrary encryption decryption using the concept of confusion and diffusion.

The concept of cloud multi-tenancy has sparked research into how to locate a specific virtual machine (VM) in a public cloud. It is possible for malicious users to identify the location of a targeted VM within a large-scale cluster and the proof for it can be found in [\cite{b8}. This study was extended by Xu et al. \cite{b9} and Herzberg et al. \cite{b10}, which prompted cloud providers to change their naming conventions to reduce the effectiveness of co-residency attacks based on network topology. Varadarajan built on this work by evaluating the susceptibility of three cloud providers to VM placement attacks and found that techniques such as virtual private clouds (VPC) could render some of these attacks ineffective.
\section{PROPOSED METHODOLOGY:}

While investigators can share their data simply through giving login
information to collaborators, a more systematic approach for sharing
data across research consortia or networks can provide more
flexibility and benefits. SciPort comes with comprehensive data
sharing capabilities: i) Convenience: data sharing is performed by a
single action and data can be selectively shared; ii) Ownership of
data: researchers own and manage their data by their own;  iii)
Flexible sharing control: data sharing can also be revoked by
researchers at any time;  iv) Up-to-date of shared data. As data are
updated or removed, corresponding shared data also need to be
synchronized accordingly to stay current;  v) Consistently
aggregated shared data from distributed sites; and vi) Lightweight.
Sharing is manipulated through metadata, no copy of large volume
data is needed.

These sharing capabilities are implemented through a lightweight,
central server based approach, as discussed next.

There would be four stages in which VM allocation happens as depicted in above figure. Their N VMs are monitored, and their performance is measured, and their anomaly scores are computed using \cite{b11}. Then these scores are passed to multi-arm bandit algorithms and VM allocation takes place. Note here that anomaly detection is only needed for training and calibration.

We want to predict a sequence of VM allocations, where our only information about the future VM comes from recommendations of a set of past recommendations. One of the major issues is that VM might be unreliable, even malicious. In such a noisy environment, we expect the most reliable VM to be allotted. Initially there is no prior information about which VM is expected to be allotted at different demand cycles, so once the VM is allotted randomly its score being malicious is observed at different allocation cycles. Now once we have more information about VM, one needs to decide should we allot the VM which has less anomaly score, or that VM having less information about- not allotted or rarely allotted. A multi-armed bandit algorithm is used to learn an optimal balance for allocating VMs between a fixed number of VM in a situation such as rare access to anomaly score by learning an efficient explore vs. exploit policy.

This is our goal for the multi-arm bandit problem and having such a strategy would prove very useful in VM allocation policy where one would like to select best VM out of a group of VMs. In this article, we approach the VM allocation policy using the multi-armed bandit problem with a classical reinforcement learning technique of an epsilon-greedy agents and others with a learning framework of reward-average sampling to compute the action-value to support the agent in improving its future action decisions for long-term reward maximization. State, action, and reward are the three main stages in the RL domain. Consider that k virtual machines are up for allocation, and that you select one VM and allocate it at each stage. the outcome after VM allocation.  Such action yields the corresponding reward in each allocation cycle. Effectively a state is the current estimation of all VMs, which could be zeros for all in the beginning, the action is the VM you decide to choose at each allocation cycle, and the reward is the result after VM allocation is done. 

Epsilon Greedy and UCB 1 algorithm are used in proposed virtual machine allocation policy.

\begin{figure}[htb]
  \centering
    \includegraphics[width=\columnwidth]{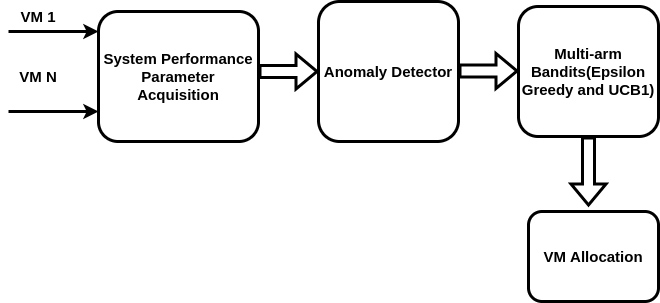}
    \caption{Proposed Virtual Machine Allocation Strategy.}
    \label{fig:Proposed Virtual Machine Allocation Strategy.}
\end{figure}

\subsection{Epsilon Greedy}
The epsilon-greedy algorithm is a technique for selecting the best performing option or "arm" at each time step, while still allowing a percentage of random selection to explore the other available options. The exploration rate, denoted by the symbol e, is an important parameter as it controls the frequency of exploring new options versus exploiting the best-known options. This algorithm has several advantages, including its ease of explanation, the simplicity in optimizing the exploration rate, and its ability to deliver good results despite its simplicity. The intuition behind this algorithm is that as more information is obtained about the available options, the need for exploration diminishes, and relying solely on exploitation becomes more effective. An alternative approach to the epsilon-greedy algorithm is the epsilon-first strategy, which involves completely random selection for a fixed number of times before switching to exploitation. Although not discussed in detail here, this strategy is worth mentioning as an option.
\subsection{UCB1}
UCB1 (Upper Confidence Bound Algorithm)
The Epsilon greedy algorithm is a widely used technique in bandit problems, but its random selection of arms can sometimes lead to inefficiencies. For instance, the algorithm may randomly select an arm that has very low performance, while ignoring an arm with significantly better performance. To address this issue, the Upper Confidence Bound (UCB) algorithms were introduced as a more efficient class of bandit algorithms.

UCB algorithms construct a confidence interval to estimate the true performance of each arm, considering the limited sample of pulls for each arm and the variance in the data. These algorithms then optimistically assume that each arm performs as well as its upper confidence bound (UCB) and select the arm with the highest UCB. This approach has several benefits, including the ability to control the trade-off between exploration and exploitation by adjusting the size of the confidence interval. This means that a larger confidence interval will lead to more exploration, while a smaller interval will favor exploitation. 

Moreover, using UCB algorithms allows for more efficient exploration of the available arms compared to the epsilon greedy algorithm. This is because the confidence intervals shrink as more data points become available, allowing the algorithm to focus on the best performing arms while still periodically exploring less explored arms with wider confidence intervals.

\subsection{Thompson Sampling}
A potent technique that can aid in resolving the exploration-exploitation conundrum in the multi-armed bandit problem is Thompson Sampling, also known as Posterior Sampling or Probability Matching. In this algorithm, actions are taken repeatedly, and this is referred to as exploration. Unlike other algorithms that provide explicit instructions for which actions to take, Thompson Sampling uses training information to evaluate the actions that are taken. This creates a need for active exploration and trial-and-error in order to find the best behavior.

After each action is taken, the machine is given a reward of 1 for a positive outcome or a penalty of 0 for a negative outcome. This information is then used to guide further actions, with the goal of maximizing the reward and improving future performance. Overall, Thompson Sampling is a sophisticated algorithm that can help to optimize decision-making in complex situations by balancing the need for exploration and exploitation.

\newpage

\section{ALGORITHM}

\begin{figure}[H]
    \centering
    \includegraphics[width=\linewidth]{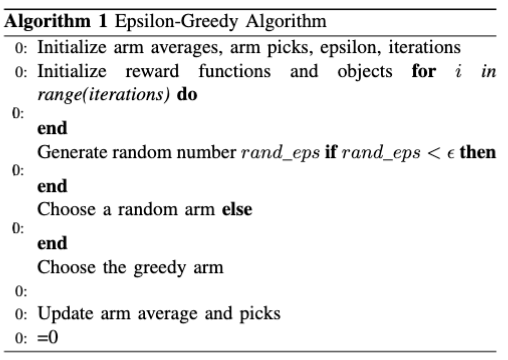}
    \caption{Description of Algorithm 1}
\end{figure}

\begin{figure}[H]
    \centering
    \includegraphics[width=\linewidth]{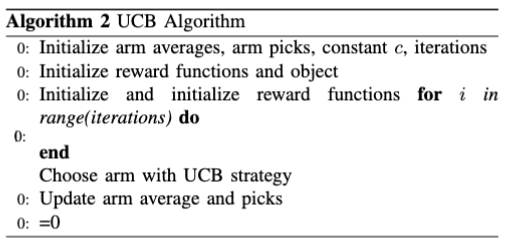}
    \caption{Description of Algorithm 2}
\end{figure}

\begin{figure}[H]
    \centering
    \includegraphics[width=\linewidth]{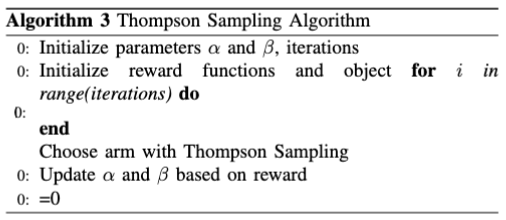}
    \caption{Description of Algorithm 3}
\end{figure}


\section{Dataset and experiment setup} 
\subsection{Dataset Generation}
We are considering 10 VMs for allocation. Based-on \cite{b11}, we generated 5000 anomaly scores for each VM at different time instances using Gaussian distribution. One anomaly score per hour, this roughly approximates 208 days.  Generally, an anomaly score of one would indicate the perfect malicious machine while zero would indicate the normally behaving machine.

\section{Dataset and Experiment Setup} 

\subsection{Dataset Generation}
We are considering 10 VMs for allocation. Based on \cite{b11}, we generated 5000 anomaly scores for each VM at different time instances using Gaussian distribution. One anomaly score per hour, this roughly approximates 208 days. Generally, an anomaly score of one would indicate the perfect malicious machine while zero would indicate the normally behaving machine.

\subsection{Results}
\begin{figure}[H]
    \centering
    \includegraphics[width=\linewidth]{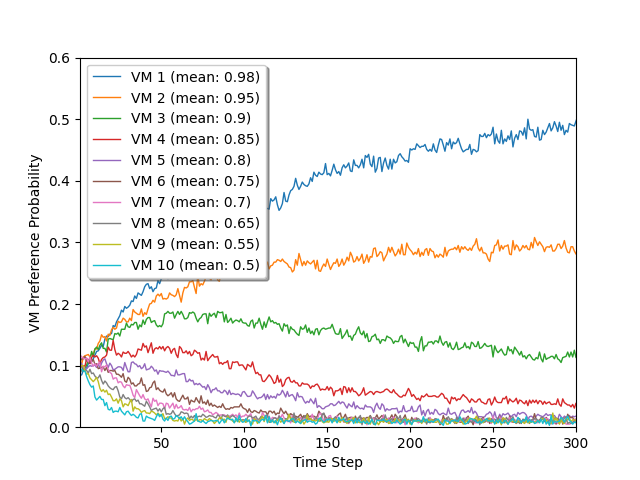}
    \caption{Virtual Machine allocation probabilities calculated per time step.}
\end{figure}

Based on the results, VM1 has the maximum chance of allocation. The average reward is also high for VM1.

\begin{figure}[H]
    \centering
    \includegraphics[width=\linewidth]{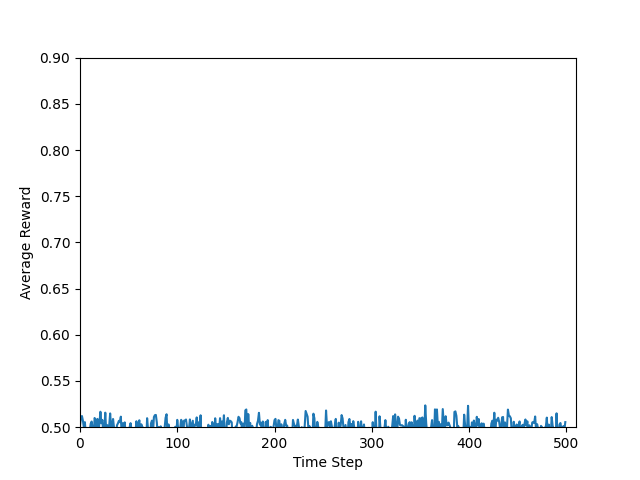}
    \caption{Rewards earned by VM with Probability per time step = 0.5}
\end{figure}

\begin{figure}[]
    \centering
    \includegraphics[width=\linewidth,height=8cm]{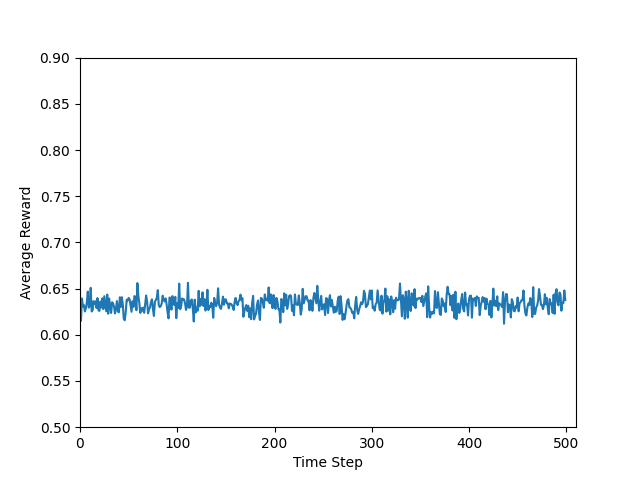}
    \caption{Rewards earned by VM with Probability per time step = 0.65}
\end{figure}

\begin{figure}[]
    \centering
    \includegraphics[width=\linewidth,height=8cm]{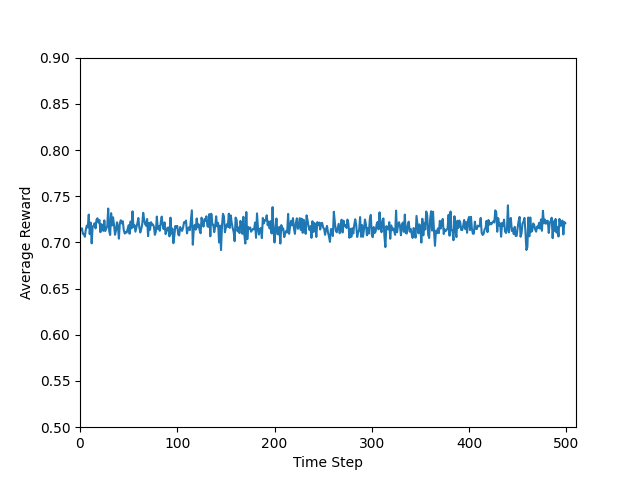}
    \caption{Rewards earned by VM with Probability per time step = 0.75}
\end{figure}

\begin{figure}[]
    \centering
    \includegraphics[width=\linewidth]{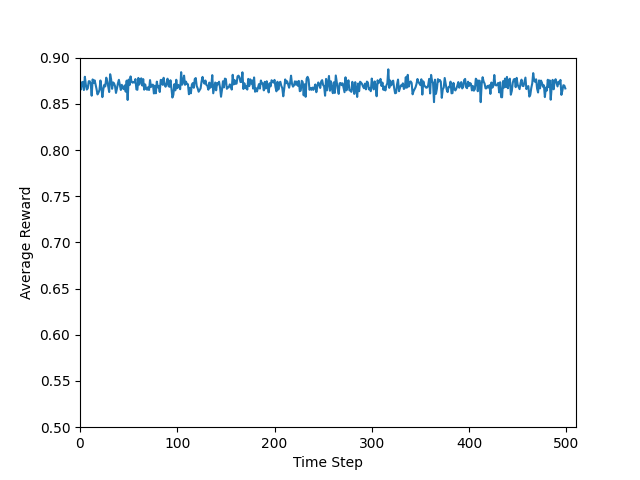}
    \caption{Rewards earned by VM with Probability per time step = 0.98}
\end{figure}

\begin{figure}[]
    \centering
    \includegraphics[width=\linewidth,height=8cm]{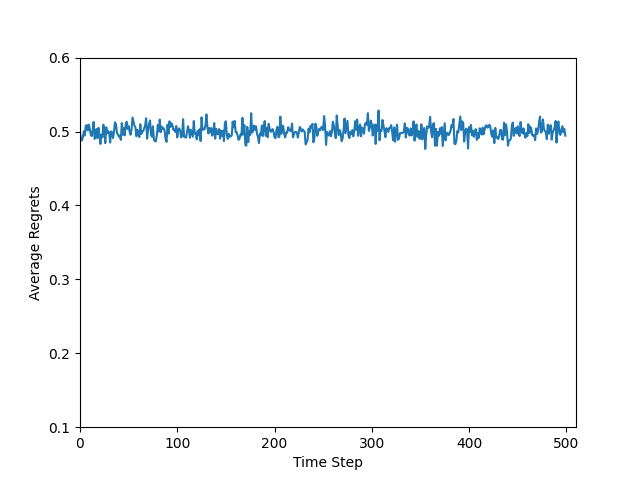}
    \caption{Regrets earned by VM with Probability per time step = 0.5}
\end{figure}

\begin{figure}[]
    \centering
    \includegraphics[width=\linewidth,height=8cm]{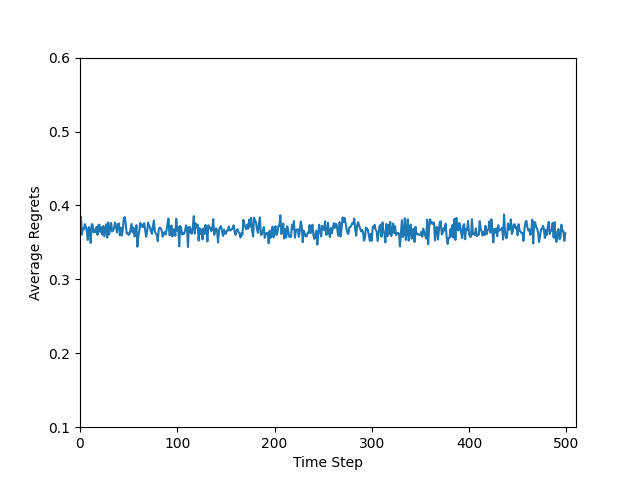}
    \caption{Regrets earned by VM with Probability per time step = 0.65}
\end{figure}

\begin{figure}[]
    \centering
    \includegraphics[width=\linewidth,height=8cm]{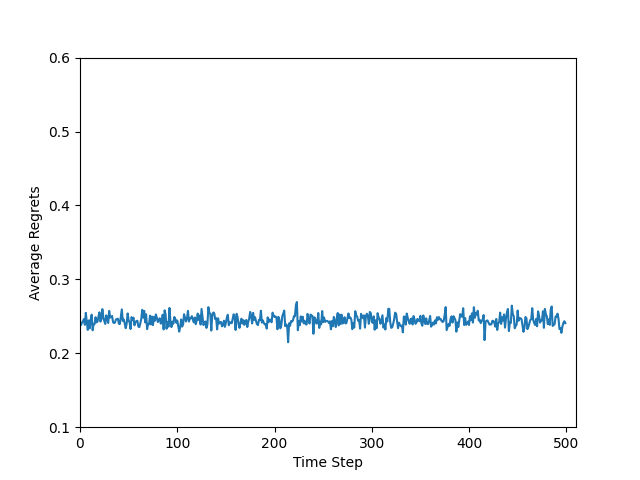}
    \caption{Regrets earned by VM with Probability per time step = 0.8}
\end{figure}

\begin{figure}[]
    \centering
    \includegraphics[width=\linewidth,height=8cm]{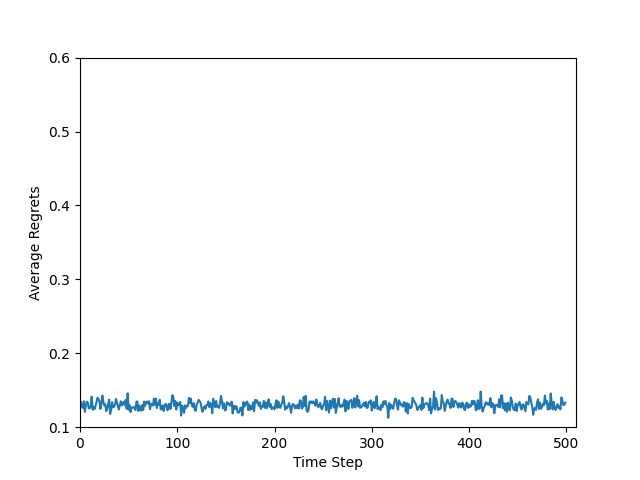}
    \caption{Regrets earned by VM with Probability per time step = 0.98}
\end{figure}
    
\begin{table}[H]
\centering
\begin{tabular}{|c|c|c|}
\hline
\textbf{VM NAME} & \textbf{Preference Probability} & \textbf{Rewards} \\
\hline
VM1 & 0.98 & 0.8699699 \\
\hline
VM2 & 0.95 & 0.8532210 \\
\hline
VM3 & 0.90 & 0.8237321 \\
\hline
VM4 & 0.85 & 0.7907654 \\
\hline
VM5 & 0.8 & 0.7555953 \\
\hline
VM6 & 0.75 & 0.7168073 \\
\hline
VM7 & 0.7 & 0.6760537 \\
\hline
VM8 & 0.65 & 0.6335874 \\
\hline
VM9 & 0.55 & 0.5451657 \\
\hline
VM10 & 0.5 & 0.4995684 \\
\hline
\end{tabular}
\caption{REWARDS EARNED BY EACH VIRTUAL MACHINE IN EACH TIME STEP IN 500-TIME STEPS.}
\end{table}

\newpage
The below plot shows epsilon greedy algorithm with different values of epsilons from 0.1 to 0.4 and UCB 1 for the best VM. The more regret algorithm has means a bad VM is to be allotted. The UCB learns better VM allocation policy and has learnt the best VM. As seen from the graph UCB has the lowest regret for VM1 and thus VM would be allocated based on the policy.
\begin{figure}[H]
\includegraphics[width=\linewidth, height= 6cm]{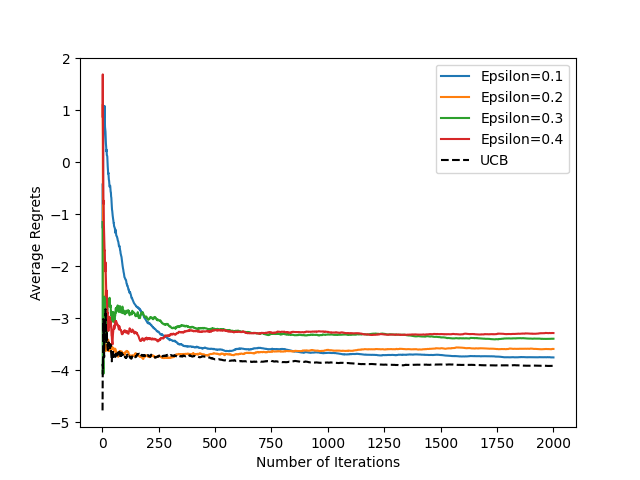}
    \caption{Regrets earned by virtual Machine using UCB vs Epsilon Greedy Method in 2000-time steps.}
\end{figure}

\begin{table}[H]
    \centering
    
    \begin{tabular}{|c|c|c|}
    \hline
         \textbf{VM NAME}& \textbf{Preference Probability} &\textbf{Regrets}\\
         \hline
         VM1 &0.98 & 0.13000300493547656\\
         \hline
         VM2 &0.95 &0.14677899267818123\\
         \hline
         VM3 &0.90 & 0.17626787165447283 \\
         \hline
         VM4 &0.85 & 0.2092324508302787 \\
         \hline
         VM5 &0.8 &  0.0.2444046877681307 \\
         \hline
         VM6 &0.75 & 0.28319260434966553 \\
         \hline
         VM7 &0.7 &  0.32394625504245034 \\
         \hline
         VM8 &0.65 & 0.36641251834004895\\
         \hline
         VM9 &0.55 & 0.4548342919317785\\
         \hline
         VM10 &0.5 & 0.5004031596862766\\
         \hline
         \end{tabular}
         
    \caption{ REGRETS EARNED BY EACH VIRTUAL MACHINE IN EACH TIME STEP IN 500-TIME STEPS.}
\end{table}

The following plot shows epsilon greedy mechanisms with different values of epsilons from 0.1 to 0.4 and Thompson sampling method for the best VM. The more regret algorithm has means a bad VM is to be allotted. For the epsilon greedy method with epsilon value 0.1, regret values are ranging from 1 to 20. For epsilon value 0.2, regret values are ranging from 1 to 12.5. For epsilon values 0.3 \& 0.4, regret values are ranging from 1 to 15 \& 1 to 14 respectively. For the Thompson sampling method, regret values are ranging from 0 to 5. Regret values of Thompson sampling method are much lesser than regret values of epsilon greedy method. The Thompson sampling method learns better VM allocation policy and has learnt the best VM. As seen from the diagram Thompson sampling method has the lowest regret for VM1 and thus VM would be allocated based on the lesser regret value.

\begin{figure}[H]
\includegraphics[width=\linewidth]{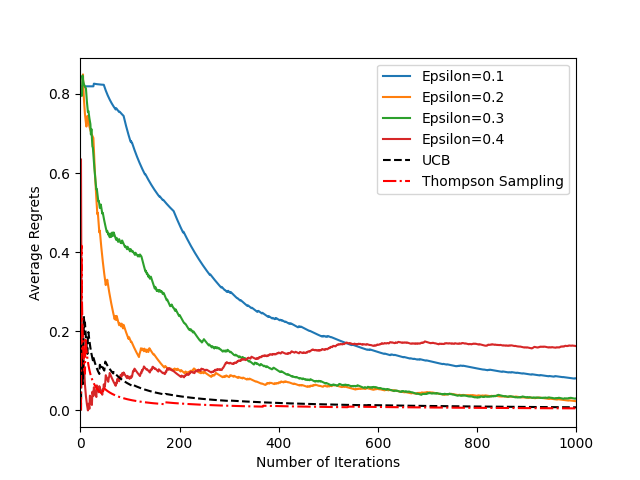}
    \caption{Rewards earned by virtual Machine using UCB vs Epsilon Greedy Method in 1000-time steps.}
\end{figure}

\end{document}